\documentclass[aps, prb, twocolumn, showpacs,amsmath,amssymb,superscriptaddress,floatfix]{revtex4-2} 
\usepackage{braket}
\usepackage{cancel}
\usepackage{titlesec}
\usepackage[utf8]{inputenc}
\usepackage[dvipdfmx]{graphicx}
\usepackage{here}
\usepackage[dvipdfmx]{color}
\usepackage{dcolumn}
\newcommand{\ketu}{\mbox{$|u\rangle $}}
\newcommand{\ketd}{\mbox{$|d\rangle $}}

\usepackage{soul}

\begin{document}

\title{Transmission of a single electron through a Berry ring} 
\author{ Maseim Bassis Kenmoe}
\email[Corresponding author: ]{kenmoe@aims.edu.gh}
\affiliation{Mesoscopy and Multilayer Structure Laboratory, Faculty of Science, Department of Physics, University of Dschang, Cameroon}
\affiliation{Institute for Theoretical Physics, University of Regensburg, Regensburg, Germany}
\author{Yosuke Kayanuma}
\email[Corresponding author: ]{kayanuma@omu.ac.jp}
\affiliation{Laboratory for Materials and Structures, Institute of Innovative Research, Tokyo Institute of Technology, 4259 Nagatsuta, Yokohama 226-8503, Japan}
\affiliation{Department of Physical Science, Osaka Prefecture University, Sakai, Osaka 599-8531, Japan}
\date{\today}
\begin{abstract}
A theoretical model of transmission and reflection of an electron with spin is proposed for a mesoscopic ring 
with rotating localized magnetic moment. This model may be realized in a pair of domain walls connecting two 
ferromagnetic domains with opposite magnetization. If the localized magnetic moment 
and the traveling spin is ferromagnetically coupled and if the localized moment rotates with opposite chirality 
in the double-path, our system is formulated in the model of an emergent spin-orbit interaction in a ring. 
The scattering problem for the transmission spectrum 
of the traveling spin is solved both in a single path and a double path model.  
In the double path, the quantum-path interference changes dramatically the transmission spectrum 
due to the effect of the Berry phase. Specifically, the spin-flip transmission and reflection are both strictly forbidden. 
\end{abstract}
\maketitle
\section{Introduction}

As an elementary particle, an electron has two basic properties, 
namely the electric charge $-e$ and the spin angular momentum $\hbar/2$ (where $\hbar$ is the reduced Planck constant). 
The quantum transport of electrons in nanostructures form a diverse field which has been intensively 
studied in recent decades\cite{Ferry}. One of the central issues in this subject is the quantum coherence, or 
the quantum-path interference observed in the transport of electrons in matter. In this connection, 
ring structures of mesoscopic scale provide us with fascinating play-grounds for quantum-path interference\cite{Formin}.
\par
As a typical example, we may name the observation of the Aharonov-Bohm (AB) effect in fabricated 
mesoscopic rings of normal metals and semiconductors \cite{Webb1985, Washburn1986} . 
The oscillation of the magnetoresistance 
with period $h/e$ was a manifestation of the AB effect\cite{Aharonov1959}, 
which proves that the vector potential linking 
the closed circuit is a real physical quantity that modulates the quantum phase of the encircling electrons. 
It should be noted that the AB effect originates from the electromagnetic interaction of the point charge 
and is independent of the spin degrees of freedom. 
\par
Another topic in this subject in recent years is the spintronics in nanostructures of magnetic materials. 
In the absence of external magnetic field, ferromagnetic materials are usually divided into mesoscopic domains 
of ordered phase. In the wall of neighboring domains with opposite magnetization, a gradual rotation of 
the localized magnetic moments connects the two domains continuously. 
\par
In 1999, Ono and coworkers\cite{Ono1999} reported, with an ingenious experimental setting, the observation of 
propagation of the magnetic domain walls in submicronmeter size wires of NiFe 
under the external magnetic field. After this, a number of experimental 
studies have been devoted to the observation and control of the motion of domain walls\cite{Harder2016}. 
With the aim to utilize 
the magnetic domain walls for the element of memory and logic devices, the studies on the fabrication and 
measurements of the nanowires of magnetic materials are accumulated\cite{Allwood2005, 
Hernandez2017} . 
\par
On the other hand, the transmission of electrons through a domain wall has long been a subject of interest because 
the interaction of electron spins and the localized magnetic moments will affect the transmission 
probability and thus plays an important role of determining the macroscopic resistance of such 
materials\cite{Cabrera1974, Korenman1977,  Berger1978, Gregg1996, Levy1997, Tatara1997, 
Gorkom1999, Tang2004}. 
Theoretically, the spin-dependent problem of transport has attracted attention not only  for domain walls 
but also in various settings of mesoscopic systems\cite{Meir1989, Loss1990, Aronov1993, Hatano2007, Fujita2011, Lobos2008}. 
The effect of Rashba spin-orbit interaction\cite{Rashba1960} in mesoscopic structures has also been studied 
for potential application to spin-interference devices\cite{Datta1990, Nitta1999, Meijer2002, Diago-Cisneros2013, Frustaglia2020, Sarkar2020}. From theoretical interest, it may also be regarded as a kind of quantum tunneling 
of a particle with internal degrees of freedom\cite{Saito1994, Penkov2000, Bertulani2007, Kilcullen2017}.


\par
In the present paper, we consider the transmission of a single electron with spin through a single domain wall 
as well as a pair of domain walls with opposite chirality of the localized magnetic moment. For simplicity, 
the whole processes are regarded as coherent processes, and we concentrate on the calculation of the transmission 
and reflection probabilities with spin-flip and spin-nonflip. The results derived here could be incorporated 
into the Landauer-B\"{u}ttiker formula\cite{Landauer1970, Buttiker1986} 
to evaluate the currents in actual experimental data. It will be shown that the spin-flip transmission through a 
coupled double-domain walls with opposite chirality is totally forbidden. Furthermore, 
the spin-conserved transmission spectrum 
shows a sharp oscillatory line-shape as a function of the incident energy. This is a result 
of the quantum-path interference due to the Berry phase\cite{Berry1984}. 
\par
In the next section, the problem is formulated as a simplified one-dimensional scattering problem for 
a single and double path model of transmission lines. 
The calculation of the transmission and reflection probabilities for the two models 
are done in Sec. III. Also, the eigenvalue problem is solved for an isolated ring in that section. The conclusions 
are given in Sec. IV.

\section{Model}
First, we consider transmission of an electron in ferromagnetic nanowires through a single domain wall as shown 
in Fig. 1(a). It is assumed that the left domain is filled with localized magnetic moment with up direction 
and the right domain with down. Actually in a very narrow ferromagnetic nanowires, it is considered that 
the magnetic moment has an easy axis parallel to the direction of the wire axis because of the shape 
anisotropy\cite{Ono1999}. For definiteness, we assume here the up and down direction for the localized 
magnetic moments in the domain. For the domain wall, there are typically two kinds, the Bloch-wall\cite{Bloch1932},
and the Neel-wall\cite{Neel1956}. In the Bloch-walls, the localized magnetic moments gradually rotate from up to down within the plane 
perpendicular to the axis connecting the two domains. In the Neel-walls, the localized moments rotate within the plane containing 
the axis which connects the domains. We assume here the Bloch-type domain wall.
\par
\begin{figure}[!t]
\centering
\includegraphics[scale=0.5]{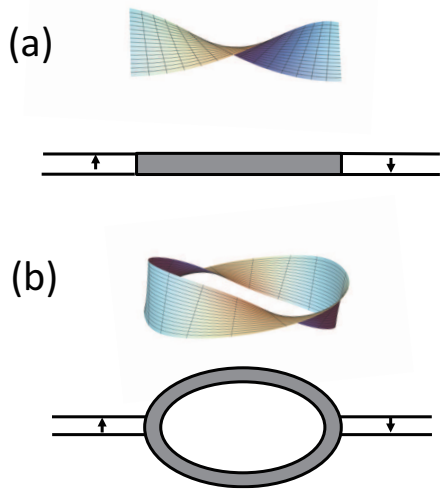}
\caption{Model of (a) a single and (b) double domain-walls in ferromagnetic nanowires. The small arrows in the 
left and right leads indicate the magnetization in the ferromagnetic regions. The shaded region represents the 
domain walls. In the upper part of (a) and (b), the rotation of the localized magnetic moments in each domain 
wall is schematically shown.} 
\label{model}
\end{figure}

It is assumed that the domain wall with length $L$ connects the left and the right domain along the $y-$axis, $0\leq y \leq  L$.
 For the direction of the localized magnetic moment and the electron spin, we introduce the $(x, z)$ axes. Note that the axes $(x,z)$ are for the internal degrees of freedom, and should not be confused with the dynamical variable $y$. 
However, it is convenient to consider a set of variables $(x, y, z)$ in order to visualize the spatial pattern of the magnetic moments as shown in  Fig.\ref{model}.
We are interested in the probability of transmission and reflection of an electron with spin, hereafter called 
a traveling spin or simply a spin, interacting with the localized magnetic moment. The wave function of the traveling 
spin is written as 
\begin{equation}
|\psi(y)\rangle=a(y)\ketu+b(y)\ketd,
\label{spin}
\end{equation}
where $\ketu$  and $\ketd$  are up-spin and the down-spin states, respectively, and $a(y)$ and $b(y)$ are 
their amplitudes. 
We consider a one-dimensional tunneling problem of the traveling spin described by the Hamiltonian 
\begin{equation}
H_0=-\frac{\hbar^2}{2m} \frac{d^2}{dy^2} +\vec{M }\cdot \vec{\sigma},
\end{equation}
in which $m$ is the effective mass of the electron, $\vec{M}$ is the localized magnetic moment, and $\vec{\sigma}$ 
are Pauli spin matrices, $\vec{\sigma}=\left(\sigma_x,\sigma_y,\sigma_z \right)$.
It is assumed that $\vec{M}$ is a smoothly varying function of $y$. See Meijer\cite{Meijer2002} for the derivation of one-dimensional 
model from a two-dimensional Hamiltonian in the case of thin nanowires. 
\par
It is assumed that, in the domain wall, the traveling spin and the localized moment $\vec{M}$ are ferromagnetically coupled with 
the energy,
\begin{equation}
\vec{M}\cdot\vec{\sigma}=-M_0\left(\cos \theta \sigma_z-\sin \theta \sigma_x\right),
\label{rotation}
\end{equation}
where $\theta$ is the angle of $\vec M$ with respect to the $z$-axis, and $M_0$ is its amplitude of interaction with 
the traveling spin. This is the $sd$-exchange interaction which is originated from quantum mechanical 
scattering processes\cite{Tatara1997}, and should not be confused with the classical electromagnetic interaction between 
the magnetic moment of the traveling spin  
and the magnetic field in the ferromagnetic material. 
\par
We may assign the chirality of the localized 
moment in the Bloch wall. It is defined as the direction of rotation of $\vec{M}$ viewed from behind along 
the $y$-axis. We assign the chirality $+$ to the domain wall in which the localized moment rotates 
in the counter-clock wise as shown in Eq. (\ref{rotation}) and in Fig.1 (a). 
It should be noted that 
Bloch walls with chirality $-$ are also possible in which the magnetic interaction is written as 
\begin{equation}
\vec{M}\cdot\vec{\sigma}=-M_0\left(\cos \theta \sigma_z+\sin \theta \sigma_x\right).
\label{rotation2}
\end{equation}
In the left domain $y<0$, 
the interaction is set to be 
$$ \vec{M}\cdot \vec{\sigma}=-M_0\sigma_z,$$ 
and in the right domain, 
$$ \vec{M}\cdot \vec{\sigma}=M_0\sigma_z.$$
\par
In actual cases, the magnetic interactions with the traveling spin may induce the change of the direction of $\vec{M}$. 
This will pose an interesting problem of the movement of the domain wall caused by the current\cite{Berger1978, Yamaguchi2004,Tatara2004}. 
We assume, in the present paper, that the localized moments are \textit{heavy} enough or intrinsically 
pinned\cite{Koyama2011} 
so that their direction is unaffected by the back-reaction from the traveling spin.  The localized magnetic moments 
are considered to be a source of static potential to the traveling spin.
\par

\begin{figure}
\centering
\includegraphics[scale = 0.55]{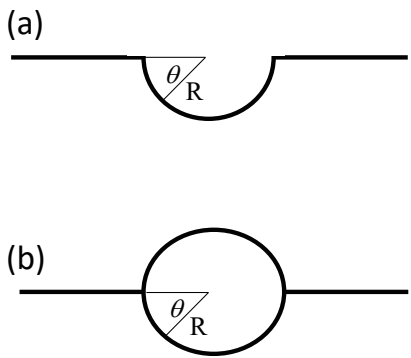}
\caption{Model of transmission lines for (a) a single domain wall and (b) double domain wall. } 
\label{transmissionline}
\end{figure}
Furthermore, the angle $\theta$ of the localized moment is assumed to change as a function of $y$ as 
$\theta (y)=(y/L)\pi$ for $0\leq y\leq L$. It is convenient to extend the definition of $\theta$ to 
$-\infty <  \theta < \infty$. By introducing the unit of energy $E_0=\hbar^2\pi^2/(2m L^2)$, the Hamiltonian $H_0$ is 
rewritten into a very simple dimensionless form, $H_s=H_0/E_0$ as 
\begin{equation}
H_s=-\frac{d^2}{d\theta^2}+V_s(\theta).
\end{equation}
Here, the interaction $V_s(\theta)$ is given by
\begin{eqnarray}
V_s(\theta)&=&-\mu\sigma_z,\quad \theta < 0, \nonumber \\
&=&-\mu\left(\cos \theta \sigma_z-\sin \theta \sigma_x\right ), \quad 0\leq \theta \leq \pi \nonumber \\
&=& \mu \sigma_z, \quad \pi< \theta,
\label{Singlepathpotential}
\end{eqnarray}
with $\mu=M_0/E_0$ being the normalized magnetization. The magnitude of $\mu$ depends both on the magnetic 
interaction $M_0$ and the width of the wall. Note that with this definition of the variable $\theta$, 
both the position of the traveling spin and its interaction energy are expressed simultaneously by $\theta$. 

\par
The Pauli matrices are written as 
\begin{eqnarray}
\sigma_x&=&|u\rangle\langle d|+|d\rangle\langle u|, \nonumber\\
\sigma_y&=&-i\left(|u\rangle\langle d|-|d\rangle\langle u|\right),\nonumber\\
\sigma_z&=&|u\rangle\langle u|-|d\rangle\langle d|.
\end{eqnarray}
In the two-vector representation of $\ketu$ and $\ketd$, they are explicitly written as
\begin{equation}
\sigma_x=
\begin{pmatrix}
          0& 1\\
          1& 0 \\
\end{pmatrix},
\quad
\sigma_y=
\begin{pmatrix}
          0& -i\\
          i& 0 \\
\end{pmatrix},
\quad
\sigma_z=
\begin{pmatrix}
          1& 0\\
          0& -1 \\
\end{pmatrix},
\nonumber
\end{equation}
and satisfy the well-known relations,
\begin{equation}
\sigma_x\sigma_y=i\sigma_z, \quad \sigma_z\sigma_x=i\sigma_y,\quad  \sigma_y\sigma_z=i\sigma_x. 
\end{equation}

\par
Next, we will consider a case of transmission of a spin through a pair of domain walls. It is assumed that these two 
domain walls have an identical width and are isolated from each other by a strip of nonmagnetic material or a void. 
This type of nanostructures will be fabricated by connecting two identical ferromagnetic nanowires to 
two common leads, and by introducing domain walls to the two bridges. Although this is a challenge, 
it will be achieved by the states of the art fabrication technique of magnetic domain conduit\cite{ Hernandez2017}.  
\par
We are 
interested in the interference of two paths of transmission of a spin. The spacial pattern of localized moments in 
the domain wall may have two choices with respect to the chirality, either it is both the same or opposite each other. 
Since the case of the same chirality is trivial, we concentrate here on the case where the chirality is opposite, namely 
localized moments rotate in the clockwise direction and anti-clockwise direction in the respective domain walls as 
schematically shown in Fig. 1 (b). 
\par
Our model is clearly visualized by the transmission line of a mesoscopic wire and a ring connected to it as   
shown in Fig. \ref{transmissionline} (b). Note that these figures represent only the topological equivalence 
to the configuration shown in Fig. \ref{model} (b). 
\par
Fig. \ref{transmissionline} (b) reminds us of the well-known setting of the quantum transport through 
the Aharonov-Bohm ring (AB-ring)\cite{Webb1985}. In the case of AB-ring, 
an essentially spinless 
charged particle feels the gauge field due to the magnetic flux which is not in contact with the particle. 
This induces a nontrivial structure in the transmission amplitude and persistent currents 
through a mesoscopic ring of normal metals\cite{Washburn1986}. 
In contrast, we study in the present paper the case of  a particle with spin but essentially no charge 
under the contact interaction 
with the localized magnetic moments. It is expected that the Berry phase\cite{Berry1984} will play 
an essential role in determining 
the transmission spectrum. In the next section, we solve the scattering problem of a spin through a single-path and 
a double-path transmission lines, and see how the quantum interference in the Berry ring changes 
the spectra dramatically.
\par

\section{Calculations}
\subsection{Single path transmission}
First, we solve the scattering problem through a single path described by $H_s(\theta)$ for an incident electron 
with majority spin, namely up-spin, coming from the left domain. In order to achieve this goal, 
the eigenvalue problem for $H_s(\theta)$ 
in the domain wall is 
solved. The eigenstate $|\psi(\theta)\rangle$ is written as
\begin{equation}
|\psi(\theta)\rangle=a(\theta)\ketu +b(\theta)\ketd=\left(
\begin{array}{l}
a(\theta)\\
b(\theta)
\end{array}
\right) ,
\end{equation}
in a 2-vector, and we have
\begin{equation}
-\frac{d^2}{d\theta^2}\left(
\begin{array}{r}
a\\
b
\end{array}
\right) 
-\mu
\begin{pmatrix}
          \cos\theta & -\sin\theta \\
         -\sin\theta & -\cos\theta \\
 \end{pmatrix}
 \left(
\begin{array}{r}
a\\
b
\end{array}
\right)
=E
\left(
\begin{array}{l}
a\\
b
\end{array}
\right) ,
\end{equation}
for the eigenvalue $E$.
\par
In order to solve the eigenvalue problem in the domain wall, we introduce an $SU(2)$ gauge field with 
the unitary transformation\cite{Korenman1977},
\begin{eqnarray}
\tilde H_s(\theta)&=&U(\theta)H_s(\theta)U^\dagger(\theta),\\
|\tilde\psi(\theta)\rangle&=&U(\theta)|\psi(\theta)\rangle,
\end{eqnarray}
where
\begin{equation}
U(\theta)=\exp\left[{-i\frac{\theta}{2}\sigma_y}\right]=\cos\frac{\theta}{2}\mathbf{1}-i\sin\frac{\theta}{2}\sigma_y,
\end{equation}
and where $\mathbf{1}$ indicates the $2\times 2$ identity matrix. 
The transformed Hamiltonian for $0\leq \theta \leq \pi$ is written as
\begin{equation}
\tilde H_s(\theta)=-\frac{d^2}{d\theta^2}-i\sigma_y\frac{d}{d\theta}+\frac{1}{4}-\mu\sigma_z.
\label{newframe}
\end{equation}
This unitary transformation induces the change of coordinate system from the space-fixed frame to 
that rotating along the direction  of the instantaneous magnetic moment.  
If the coupling constant $\mu$ is much larger than unity, the spin-flipping term proportional to $\sigma_y$ in 
the right hand side of Eq.(\ref{newframe}) may be negligible, so that the traveling spin will follow the localized 
magnetic moment adiabatically. We call this new frame of coordinate a rotating coordinate. It should be noted that 
the second term of Eq.(\ref{newframe}) is given as a product of momentum and the spin variable. This is 
regarded as representing a kind of emergent spin-orbit interaction introduced by a gauge transformation with $U(\theta)$.
\par
Since $\tilde H_s(\theta)$ does not contain the variable $\theta$ in the potential term, we can solve the eigenvalue 
equation
$$
\tilde H_s(\theta)|\tilde\psi(\theta)\rangle = E|\tilde\psi(\theta)\rangle
$$
by the method of characteristics. Putting 
\begin{equation}
|\tilde \psi(\theta)\rangle =e^{iq\theta}\left(a \ketu+b \ketd \right),
\end{equation}
for constants $a$ and $b$ and for complex $q$, we find
\begin{eqnarray}
\left(q^2-\mu\right) a-iqb &=&\left( E-\frac{1}{4}\right) a,\nonumber\\
iqa+ \left(q^2+\mu\right) b &=& \left(E-\frac{1}{4}\right) b,
\label{eigenvalueeq}
\end{eqnarray}
which leads to the eigenvalue equation,
\begin{equation}
\left(E-q^2-\frac{1}{4}\right)^2-u^2-q^2=0.
\end{equation}
The above equation has solutions,
\begin{equation}
E=E_{\pm}(q,\mu)=q^2+\frac{1}{4}\pm \sqrt{q^2+\mu^2}.
\label{Eofq}
\end{equation}
\begin{figure}
\centering
\includegraphics[scale = 0.4]{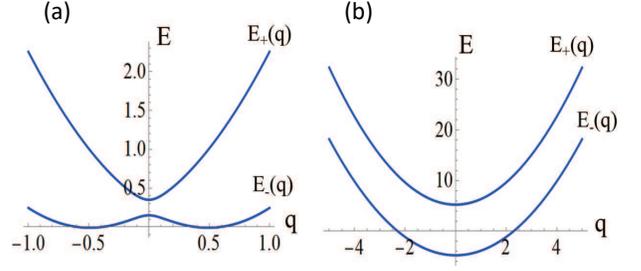}
\caption{Dispersion relation for the energy as a function of $q$ in the domain wall with 
(a) $\mu=0.1$ and (b) $\mu=5$.} 
\label{dispersion}
\end{figure}
\par
\begin{figure}[!h]
\centering
\includegraphics[scale = 0.6]{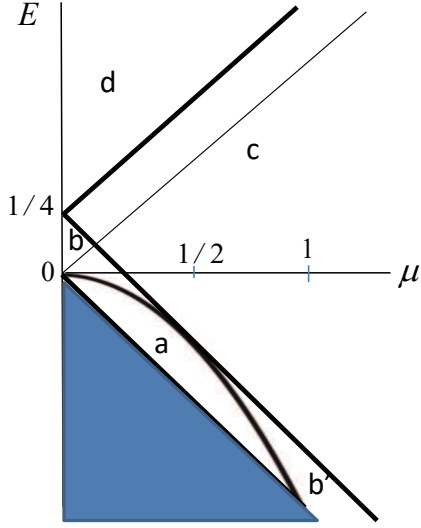}
\caption{Phase diagram for the values of $q$ in the $(E,\mu)$ plane. The dispersion relation $E=E(q,\mu)$ has 
solutions $q$ with (a) four complex numbers, (b) four real numbers, (b') four pure imaginary numbers, (c) two 
real and two pure imaginary numbers, (d) two real positive and two real negative numbers. The shaded area  
is unphysical region.} 
\label{phase}
\end{figure}

The upper branch $E=E_+(q)$ always has a minimum at $q=0$, while the point $q=0$ is a minimum point in 
in the lower branch $E=E_-(q)$ for the strong coupling $\mu>\frac{1}{4}$, but 
it becomes a local maximum point for weak coupling $\mu\leq \frac{1}{4}$. In Fig. \ref{dispersion}, 
two examples of the dispersion curves are shown. 
Inversely, for a fixed value of energy $E$, the eigenvalues of the wave number $q$ are obtained  
by solving the equation (\ref{Eofq}). 
This gives a quartet of generally complex numbers $q=\pm q_r\pm iq_i$, where $q_r$ and $q_i$ are real numbers. 
With an inspection of the solution of the quadratic equation of $q^2$, the \textit{phase diagram} representing the nature of $q$ 
is obtained as plotted in Fig. \ref{phase}.  The $(\mu, E)$ plane is divided into five regions according to the character 
of $q$, namely, (a) $-\mu\leq E\leq -\mu^2$, four complex numbers, (b) $-\mu^2\leq E\leq \frac{1}{4}-\mu$, 
and $\mu<\frac{1}{2}$, four real numbers, (b')  $-\mu^2\leq E\leq \frac{1}{4}-\mu$ and $\frac{1}{2}\leq \mu$, 
four pure imaginary numbers, (c) $\frac{1}{4}-\mu\leq E<\frac{1}{4}+\mu$, two real and two pure imaginary 
numbers, and (d) $\frac{1}{4}+\mu\leq E$, two real positive numbers and two real negative numbers. 
\par
It should be noted that the spin-conserved transmission becomes allowed energetically for $\mu<E$, with the boundary 
shown by a thin straight line in the region c. 
In the region satisfying the conditions, $\mu \leq E < \mu+\frac{1}{4}$ and $-\mu+\frac{1}{4} \leq E$, 
there are two evanescent states in addition to two propagating states in the domain wall, even though both the up-spin 
and down-spin states are energetically allowed to exist here.

\begin{figure}
\centering
\includegraphics[scale = 0.6]{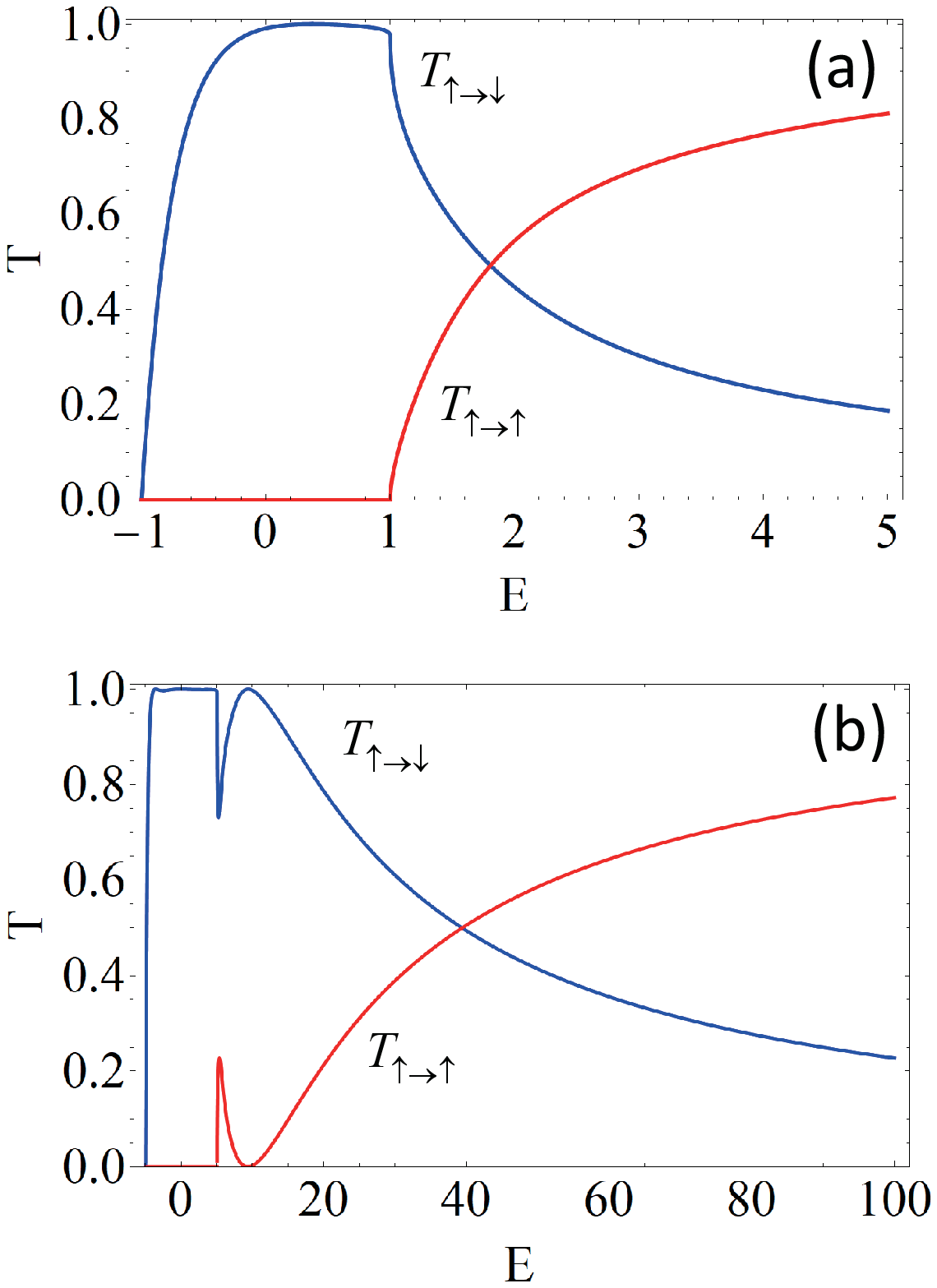}
\caption{Transmission spectra in a single domain wall with (a) $\mu=1$ and (b) $\mu=5$.} 
\label{SingleWeak}
\end{figure}
We solve the scattering problem of the traveling spin under the condition that an up-spin comes from 
the left domain with 
energy $E (\geq -\mu)$ and momentum $q_0=\sqrt{E+\mu}$. Note that the direction of the magnetic field is reversed from up in the left domain to down in the right domain. 
Therefore, the spin-conserved transmission $\ketu \rightarrow \ketu$ is allowed energetically only for 
$\mu\leq E$, while the spin-flip transmission $\ketu \rightarrow \ketd$ is allowed for $-\mu\leq E$. 
For a given incident energy $E$, the four eigenstates in the domain wall are written as 
$|\tilde\psi_1(\theta)\rangle,  |\tilde\psi_2(\theta)\rangle, |\tilde\psi_3(\theta)\rangle, |\tilde\psi_4(\theta)\rangle,$ 
where 
\begin{equation}
|\tilde\psi_j(\theta)\rangle = e^{iq_j\theta}\left( a_j\ketu+b_j \ketd \right), \quad j=1,2,3,4,
\end{equation}
in which $a_j$ and $b_j$ are determined by Eq.(\ref{eigenvalueeq}) as 
\begin{eqnarray}
a_j&=&iq_j,\\
b_j&=&-\left(\epsilon_j+\mu\right),\quad j=1,2,3,4,
\end{eqnarray}
with 
\begin{equation}
\epsilon_j=E-q_j^2-\frac{1}{4},\quad j=1,2,3,4. 
\end{equation}
\par
\begin{widetext}
The eigenfunction in the space-fixed coordinate is given by
\begin{eqnarray}
|\psi_j(\theta)\rangle &=&U^\dagger (\theta)|\tilde\psi_j(\theta)\rangle\nonumber\\
&=&e^{iq_j\theta}\Bigl\{\left(a_j\cos\frac{\theta}{2}+b_j\sin\frac{\theta}{2}\right)\ketu+
\left(- a_j\sin\frac{\theta}{2}+b_j\cos\frac{\theta}{2}\right)\ketd\Bigr\},
\end{eqnarray}

\end{widetext}
where we used the relations
\begin{eqnarray}
e^{i\frac{\theta}{2}\sigma_y}\ketu &= &\cos\frac{\theta}{2}\ketu -\sin\frac{\theta}{2}\ketd,\nonumber\\
e^{i\frac{\theta}{2}\sigma_y}\ketd &= &\sin\frac{\theta}{2}\ketu +\cos\frac{\theta}{2}\ketd.
\end{eqnarray}
The wave function $|\Psi(\theta)\rangle$ in the domain wall is then expressed as 
\begin{equation}
|\Psi(\theta)\rangle=\sum_{j=1}^4 \gamma_j|\psi_j(\theta)\rangle,
\end{equation}
where $\gamma_j$ are unknown parameters to be determined. 
\par
There are two cases according to the incident energy; In the case 1 where $\mu\leq E$, both of the spin conserved 
and spin flip transmission and reflection are allowed energetically. 
In the case 2 where $-\mu\leq E<\mu$, only the spin flip transmission and spin conserved reflection are allowed. 
\par
To begin with, we discuss the case 1. The wave function $|\Psi_{l}(\theta)\rangle$ in the left domain $(\theta<0)$
has the form 
\begin{equation}
|\Psi_{l}(\theta)\rangle=\left(e^{iq_0\theta}+re^{-iq_0\theta} \right)\ketu+\rho e^{-ik_0\theta}\ketd ,
\label{left}
\end{equation}
in which $r$ is the  amplitude of spin conserved reflection and $\rho$ is the amplitude of  
the spin flip reflection, and $q$ and $k$ are given by
\begin{eqnarray}
q_0&=&\sqrt{E+\mu},\\
k_0&=&\sqrt{E-\mu}.
\end{eqnarray}
In the right domain $(\pi<\theta)$, the wave function $|\Psi_{r}(\theta)\rangle$ is given as
\begin{equation}
|\Psi_{r}(\theta)\rangle=te^{iq_0(\theta-\pi)}\ketd +\eta e^{ik_0(\theta-\pi)}\ketu,
\label{right}
\end{equation}
in which $t$ is the  amplitude of spin flip transmission and $\eta$ is the amplitude of spin conserved transmission. 
The unknown 8 parameters $(\gamma_1,\gamma_2,\gamma_3,\gamma_4, r,t,\rho,\eta)$ are determined by the continuity 
condition of $|\Psi_i(\theta\rangle$ and their derivatives at the boundaries, $\theta=0$ and $\theta=\pi$, 
for up-spin and down-spin components. These give 8 linear simultaneous equations for 8 unknowns, which are solved 
by the inversion of matrices. 
The probability of spin-conserved transmission $T_{\uparrow\rightarrow\uparrow}$ and 
spin-flip reflection $R_{\uparrow\rightarrow\downarrow}$ are given by 
\begin{eqnarray}
T_{\uparrow\rightarrow\uparrow}&=&\frac{k_0}{q_0}|\eta|^2,\\
R_{\uparrow\rightarrow\downarrow}&=&\frac{k_0}{q_0}|\rho|^2,
\end{eqnarray}
respectively. 
The probability of spin-flip transmission $T_{\uparrow\rightarrow\downarrow}$ and 
spin-conserved reflection $R_{\uparrow\rightarrow\uparrow}$ are given by 
\begin{eqnarray}
T_{\uparrow\rightarrow\downarrow}&=&|t|^2,\\
R_{\uparrow\rightarrow\uparrow}&=&|r|^2,
\end{eqnarray}
respectively. For the case 2 below threshold $-\mu\leq E<\mu$, the same argument is proceeded 
with the only change that the 
propagating  states become evanescent states so that $k_0$ is replaced by $-i\kappa_0$ with $\kappa_0=\sqrt{\mu-E}$. 

\par
In Fig. \ref{SingleWeak}, two examples of the calculated transmission probabilities 
$T_{\uparrow\rightarrow\uparrow}$ and $T_{\uparrow\rightarrow\downarrow}$ through a single domain wall 
are presented. In Fig. \ref{SingleWeak} (a) the results for relatively weak coupling case $\mu=1$, and 
in Fig. \ref{SingleWeak} (b) an intermediate coupling case $\mu=5$ are shown. The reflection probabilities 
are small and not shown here. Note a sharp structure in $T_{\uparrow\rightarrow\uparrow}$ just above 
the threshold energy $E=\mu$ in Fig.\ref{SingleWeak} (b). This is a resonant transmission due to the existence of evanescent states 
in the domain wall, as discussed above.

\subsection{Isolated Berry ring}
We proceed to our main target; the study of transmission and reflection of a spin through double path domain 
walls, or a transmission line of Berry ring. First, we solve the eigenvalue problem for 
an isolated ring, where the radius $R$ and the polar angle $\theta$ are defined as shown in 
Fig \ref{transmissionline}. 
The width $L$ of the domain wall discussed in the Subsection A is given by $L=\pi R$. The Hamiltonian $H_r$ for the isolated ring is given by 
\begin{equation}
H_r=-\frac{d^2}{d\theta^2} +V_r(\theta).
\end{equation}
Here, the interaction potential $V_r(\theta)$ is given by
\begin{equation}
V_r(\theta)=-\mu(\cos\theta\sigma_z-\sin\theta \sigma_x), \quad 0\leq \theta \leq 2\pi.
\end{equation}
By the unitary transformation $U(\theta)=e^{-i\frac{\theta}{2}\sigma_y}$, the Hamiltonian is transformed into 
\begin{equation}
\tilde H_r(\theta) =-\left(\frac{d}{d\theta}+\frac{i}{2}\sigma_y\right)^2-\mu\sigma_z,
\end{equation}
as before. The eigenvalues are given by
\begin{equation}
E_\pm(q,\mu)=q^2+\frac{1}{4} \pm \sqrt{\mu^2+q^2},
\end{equation}
with the eigenfunctions in the rotating coordinate,
\begin{equation}
|\tilde\psi_{\pm,q}(\theta)>=e^{iq\theta}(a_{\pm,q}|u>+b_{\pm,q}|d>),
\end{equation}
in which
\begin{eqnarray}
\label{a1}
a_{\pm,q} &=&C_{\pm,q}iq,
\\
\label{b1}
b_{\pm,q}&=&-C_{\pm,q}\bigl(\mu\pm\sqrt{\mu^2+q^2}\bigr),
\end{eqnarray}
where $C_{\pm,q}$ is the normalization constant. The space-fixed coordinate representation is given as 
\begin{widetext}
\begin{eqnarray}
|\psi_{\pm,q}(\theta)\rangle&=&U^\dagger(\theta)|\tilde \psi_{\pm,q}(\theta)\rangle\nonumber\\
&=&e^{iq\theta}\Bigl\{\bigl(a_{\pm,q}\cos\frac{\theta}{2}+b_{\pm,q}\sin\frac{\theta}{2}\bigr)\ketu+
\bigl(-a_{\pm,q}\sin\frac{\theta}{2}+b_{\pm,q}\cos\frac{\theta}{2}\bigr)\ketd\Bigr\}.
\label{cycle}
\end{eqnarray}
\end{widetext}
\par
From  Eq.(\ref{cycle}), we find 
\begin{equation}
|\psi_{\pm,q}(2\pi)\rangle=-e^{i2\pi q}|\psi_{\pm,q}(0)\rangle.
\label{condition1}
\end{equation}
The factor $-1$ is nothing but the Berry phase factor, while the factor $e^{i2\pi q}$ is 
the dynamical phase factor. 
The eigenvalue of $q$ is determined by the continuity condition,
\begin{equation}
|\psi_{\pm,q}(0)\rangle=|\psi_{\pm,q}(2\pi)\rangle.
\label{condition2}
\end{equation}
From Eq.(\ref{condition1}) and (\ref{condition2}), the allowed values of $q$ are quantized as 
\begin{equation}
q=\pm\bigl(n+\frac{1}{2}\bigr),\quad n=0,1,2,\cdots.
\end{equation}
The quantum number $q$ determines the wave form of rotational motion of the electron. Each eigenstate is 
two-fold degenerate according to the sign of $q$. This is a result of the time-reversal symmetry of our 
Hamiltonian. 
\par
\begin{figure}
\centering
\includegraphics[scale = 0.5]{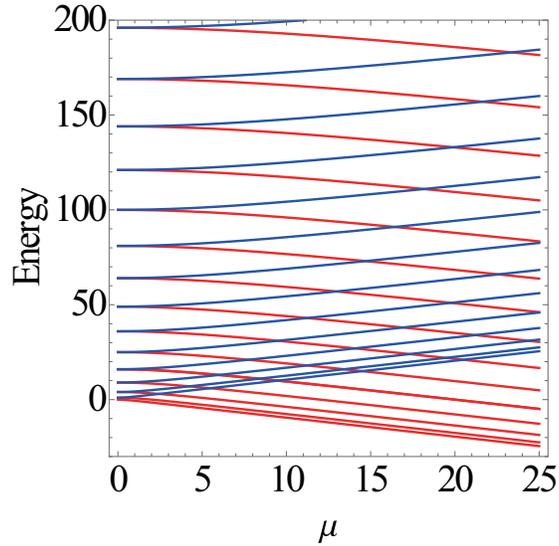}
\caption{Eigenenergies in an isolated Berry ring as a function of $\mu$. The red lines are lower branch 
$E_-(q,\mu)$ and the blue lines are the upper branch $E_+(q,\mu)$. The parameter $q$ is an orbital angular 
momentum which takes quantized values $\pm (n+\frac{1}{2}),\quad n=0,1,2,\cdots$.} 
\label{Epermu}
\end{figure}

In Fig.\ref{Epermu}, the $\mu$ dependence of the eigenenergies is plotted. The red lines correspond 
to the lower branch 
and the blue ones to the upper branch. Each level is composed of the two-fold degenerate levels 
corresponding to the signs of 
$q=\pm\bigl(n+\frac{1}{2}\bigr)$. The quantum number $q$ may be regarded as an orbital angular momentum 
in the mesoscopic ring. 
A curious thing is that it takes half-integer values to compensate the Berry phase due to the spin part. 
\par
Note that, in the limit  $\mu\rightarrow 0$, the eigenenergies $E_\pm (q,\mu)$ in Eq.(\ref{Eofq}) tend to distinct 
values $E_+(q,0)=(n+1)^2$ and $E_-(q,0)=n^2$ for $q=n+\frac{1}{2}$, and  $E_+(q,0)=n^2$ and $E_-(q,0)=(n+1)^2$ 
for $q=-(n+\frac{1}{2})$, $n=0,1, 2, \cdots$. Therefore, at $\mu=0$, 
the eigenstates corresponding to $n^2$ are four-fold 
degenerate for $n\geq 1$, and two-fold degenerate for $n=0$. This is natural because, at $\mu=0$, the eigenstates 
of the electron in the ring are degenerate with respect to the sign of the orbital angular momentum and the sign of 
the spin. 
\par
The expectation value of the spin direction at $\theta$ is calculated as
\begin{widetext}
\begin{eqnarray}
\langle \sigma_z \rangle&=&\langle\psi_{\pm,q}(\theta)|\sigma_z|\psi_{\pm,q}(\theta)\rangle
=\bigl(|a_{\pm,q}|^2-b_{\pm,q}^2\bigr)\cos\theta,\\
\langle \sigma_x \rangle&=&\langle\psi_{\pm,q}(\theta)|\sigma_x|\psi_{\pm,q}(\theta)\rangle
=-\bigl(|a_{\pm,q}|^2-b_{\pm,q}^2\bigr)\sin\theta.
\end{eqnarray}
\end{widetext}
We may define the chirality of the spatial pattern $\chi^s_\pm$ of the traveling spin by the sign of 
$\bigl(|a_{\pm,q}|^2-b_{\pm,q}^2\bigr)$
for each eigenstate. Inserting (\ref{a1}) and (\ref{b1}), we find $\chi^s_+<0$ and $\chi^s_->0$. 
This means that the traveling spin in the lower branch follows the spatial pattern parallel with the localized 
magnetic moment, and that in the upper branch anti-parallel to it. As the interaction $\mu$ is switched on, the degeneracy is lifted according to the relative relation of the chirality 
of the traveling spin $\chi^s$ and the chirality of the localized moment. This may be regarded as the 
emergent spin-orbit interaction.
\par
\begin{figure}
\centering
\includegraphics[scale = 0.5]{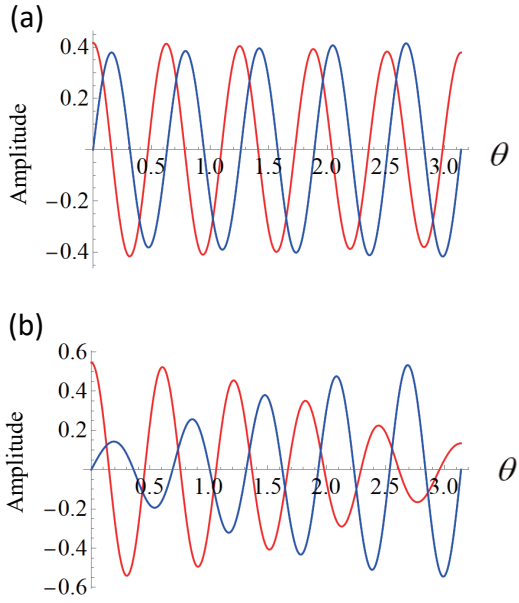}
\caption{Spin resolved amplitude of the cosine-like standing waves in the isolated Berry ring 
for (a) weak coupling case, $\mu=1$ and (b) strong coupling case $\mu=5$. The red lines are the up-spin 
amplitude and the blue lines are the down-spin amplitude. Only the half cycle $0\leq \theta\leq \pi$ is plotted here.} 
\label{spinamp}
\end{figure}

For $n\geq 0$ ($n\leq 0$), the traveling electron in the lower branch rotates 
in the counterclock wise (clock wise) direction 
in the ring with counterclock wise (clock wise) rotation of the spin. These are persistent currents. However, 
because their eigenenergies are degenerate, they form a pair of standing waves. The absence of persistent currents in 
an isolated Berry ring is a result of time-reversal symmetry. 
\par
We define a cosine-like and sine-like standing wave states for both of the lower and upper branches as
\begin{eqnarray}
|\Psi_{\pm,c}(\theta)\rangle&=&\frac{1}{2i}\bigl(|\psi_{\pm,q}(\theta)\rangle-|\psi_{\pm,-q}(\theta)\rangle\Bigr), \\
|\Psi_{\pm,s}(\theta)\rangle&=&\frac{1}{2}\bigl(|\psi_{\pm,q}(\theta)\rangle+|\psi_{\pm,-q}(\theta)\rangle\Bigr).
\end{eqnarray}
These states satisfy the conditions,
\begin{widetext}
\begin{eqnarray}
|\Psi_{\pm,c}(0)\rangle&=&|\Psi_{\pm,c}(2\pi)\rangle=C_{\pm,q}q\ketu,\quad |\Psi_{\pm,c}(\pi)\rangle
=C_{\pm,q}(-1)^n b_{\pm,q}\ketu,\label{c}\\
|\Psi_{\pm,s}(0)\rangle&=&|\Psi_{\pm,s}(2\pi)\rangle=C_{\pm,q}b_{\pm,q}\ketd,\quad |\Psi_{\pm,s}(\pi)\rangle
=C_{\pm,q}(-1)^n q\ketd.
\label{s}
\end{eqnarray}
\end{widetext}
Eqs.(\ref{c}) and (\ref{s}) imply that the spin-flip transmission through a Berry ring is totally forbidden 
if the leads are connected at $\theta=0$ and $\theta=\pi$. In Fig. \ref{spinamp}, an example of the spin profile for 
the cosine-like state $q=10+\frac{1}{2}$ is shown. As shown here, an electron confined in a Berry ring occupies 
a standing wave state with its spin winding along the ring. 
\par
\subsection{Double path transmission }
We calculate the probability of transmission and reflection of a traveling spin through a pair of domain walls by 
the simplified model of a Berry ring with attached leads as shown in Fig. \ref{transmissionline}. 
For that purpose, it is more convenient to redefine the variable $\theta$ in the top-side path as 
$2\pi-\theta\rightarrow \theta$. 
For the sake of economy of notation, we use the same variable $\theta,\quad (0\leq\theta\leq\pi)$ 
both for the top-side path of the ring and the bottom-side path of the ring measured from the left vertex. 

\par
The Hamiltonian for the double path transmission $H_d$ is written as 
\begin{equation}
H_d=-\frac{d^2}{d\theta^2}+V_d(\theta),
\end{equation}
where $V_d(\theta)$ is defined as
\begin{equation}
V_d(\theta)=-\mu\left(\cos \theta \sigma_z-\sin \theta \sigma_x\right ), \quad 0 \leq \theta \leq \pi 
\end{equation}
in the bottom-side path and
\begin{equation}
V_d(\theta)=-\mu\left(\cos \theta \sigma_z+\sin \theta \sigma_x\right ), \quad  0 \leq \theta \leq \pi 
\end{equation}
in the top-side path. 
For $\theta < 0$ and $\pi <\theta$, $V_d(\theta)$ is the same as $V_s(\theta)$,
\begin{eqnarray}
V_d(\theta)&=& -\mu\sigma_z, \quad \theta<0,\nonumber\\
               &=& \mu\sigma_z ,\quad  \pi<\theta.
\end{eqnarray}
The eigenfunctions for the bottom-side path 
$|\tilde \psi_j^b(\theta)\rangle, \quad j=1,2,3,4$ in the gauge transformed coordinate are the same as those 
$|\tilde \psi_j(\theta)\rangle$ which are obtained in the previous subsection. The eigenfunctions in the 
top-side path $|\tilde \psi_j^t(\theta)\rangle$ are given by the same argument as
\begin{equation}
|\tilde\psi_j^t(\theta)\rangle = e^{iq_j\theta}\left( a_j\ketu-b_j \ketd \right), \quad j=1,2,3,4.
\end{equation}
Note the difference in the sign before $b_j$. 
The corresponding eigenfunctions in the original space-fixed coordinate are given by

\begin{widetext}

\begin{eqnarray}
|\psi_j^b(\theta)\rangle &=&U^\dagger (\theta)|\tilde\psi_j^b(\theta)\rangle\nonumber\\
&=&e^{iq_j\theta}\Bigl\{\left(a_j\cos\frac{\theta}{2}+b_j\sin\frac{\theta}{2}\right)\ketu+
\left(- a_j\sin\frac{\theta}{2}+b_j\cos\frac{\theta}{2}\right)\ketd\Bigr\},\\
|\psi_j^t(\theta)\rangle &=&U^\dagger (-\theta)|\tilde\psi_j^t(\theta)\rangle\nonumber\\
&=&e^{iq_j\theta}\Bigl\{\left(a_j\cos\frac{\theta}{2}+b_j\sin\frac{\theta}{2}\right)\ketu+
\left(  a_j\sin\frac{\theta}{2}-b_j\cos\frac{\theta}{2}\right)\ketd\Bigr\}, \quad j=1,2,3,4,
\end{eqnarray}

\end{widetext}
respectively,
\begin{figure}
\centering
\includegraphics[scale = 0.67]{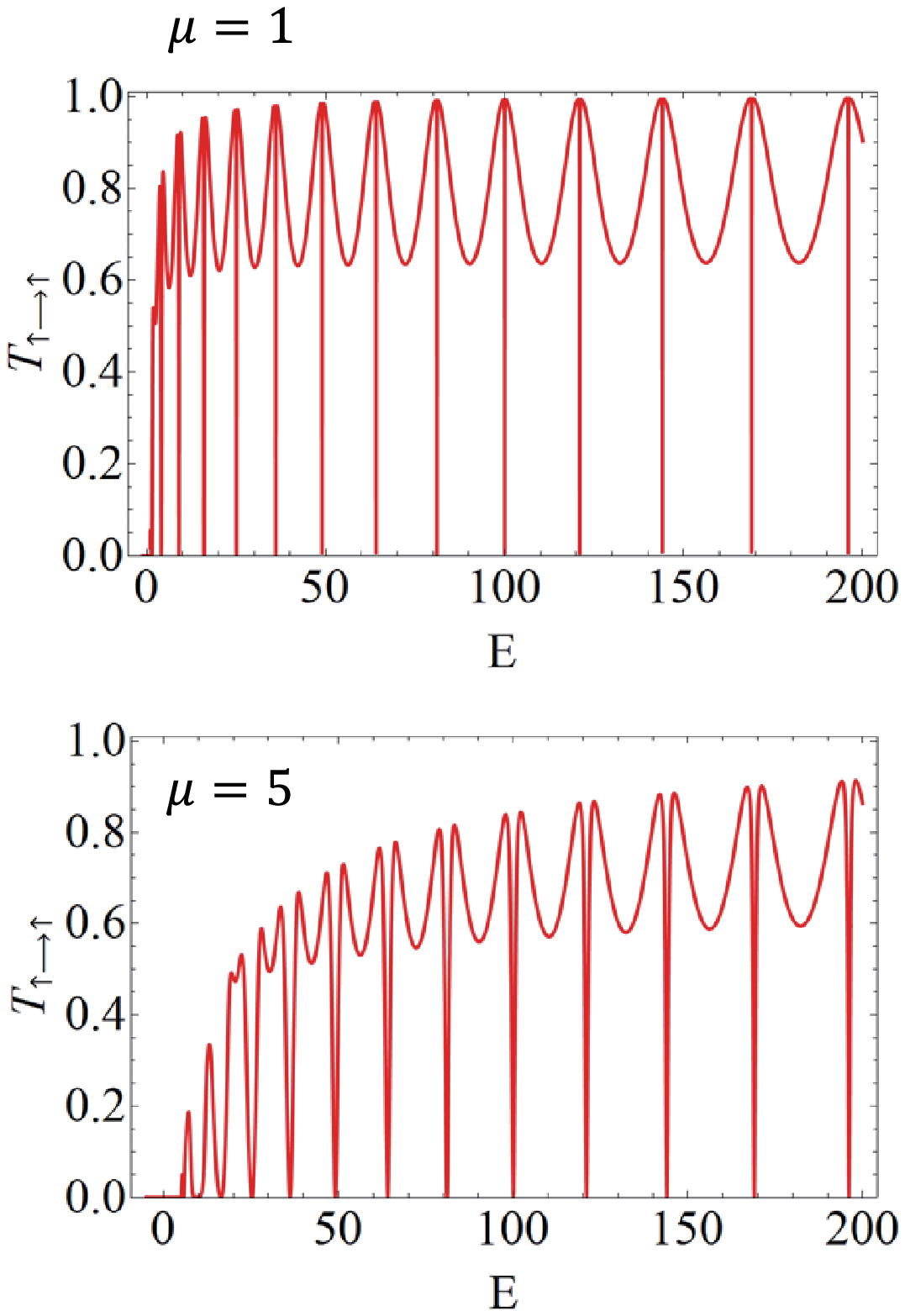}
\caption{Spin-conserved transmission spectrum for the incident up-spin in the Berry ring 
with $\mu=1$ and $\mu=5$.} 
\label{Double_weak}
\end{figure}
\begin{figure}
\centering
\includegraphics[scale = 0.6]{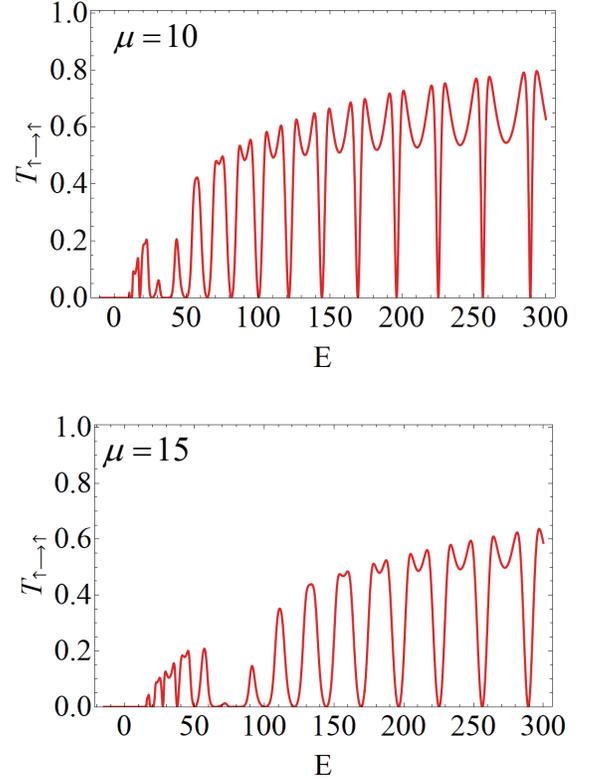}
\caption{Spin-conserved transmission spectrum for the incident up-spin in the Berry ring 
with $\mu=10$ and $\mu=15$. } 
\label{Double_strong}
\end{figure}
\par
We calculate the transmission and reflection probability of an up-spin coming from the left lead with energy 
$E(\geq \mu)$. The wave function in the left lead $|\Psi_l(\theta)\rangle$ and the right lead $|\Psi_r(\theta)\rangle$ 
are given by the same form as Eq.(\ref{left}) and Eq.(\ref{right}). 
The wavefunctions in the bottom-side path and the top-side path are given by
\begin{eqnarray}
|\Psi_b(\theta)\rangle&=&\sum_{j=1}^4 \alpha_j |\psi_j^b(\theta)\rangle,\\
|\Psi_t(\theta)\rangle&=&\sum_{j=1}^4 \beta_j |\psi_j^t(\theta)\rangle,
\end{eqnarray}
respectively. The 12 unknown quantities
 $(t,r,\rho,\eta,\alpha_1,\alpha_2,\alpha_3,\alpha_4,\beta_1,\beta_2,\beta_3,\beta_4)$ are determined by the continuity 
 conditions and the conservation of current at $\theta=0$ and $\theta=\pi$,

\begin{eqnarray}
|\Psi_l(0)\rangle&=& |\Psi_b(0)\rangle= |\Psi_t(0)\rangle,\nonumber\\
|\Psi'_l(0)\rangle&= &|\Psi'_b(0)\rangle= |\Psi'_t(0)\rangle,
\label{zero}
\end{eqnarray}
 and
\begin{eqnarray}
|\Psi_r(\pi)\rangle&=& |\Psi_b(\pi)\rangle= |\Psi_t(\pi)\rangle,\nonumber\\
|\Psi'_r(\pi)\rangle&=& |\Psi'_b(\pi)\rangle= |\Psi'_t(\pi)\rangle,
\label{pi}
\end{eqnarray}
where the dash means the derivative with respect to $\theta$. 

Eqs. (\ref{zero})  and (\ref{pi}) constitute a set of 12 conditions for the coefficients of $\ketu$ 
and $\ketd$. It is found that these simultaneous equations are greatly simplified if one  notes the symmetry 
properties of the solution. We set $A_j=\alpha_j+\beta_j$ and $B_j=\alpha_j-\beta_j$, and compare the coefficients 
of $\ketu$ and $\ketd$ on both sides of the equations. Eliminating $r,t,\rho$ and $\eta$, we find that $A_j$ and $B_j$ 
are decoupled as 
\begin{eqnarray}
\sum_j \bigl(q_j+\frac{1}{2}q_0\bigr)a_jA_j&=&2q,\\
\sum_j e^{iq_j\pi}\bigl(q_j-\frac{1}{2}k_0\bigr)b_jA_j&=&0,\\
\sum_j b_jA_j&=&0,\\
\sum_j e^{iq_j\pi}a_jA_j&=&0,
\end{eqnarray}
for $A_j$ and 
\begin{eqnarray}
\sum_j \bigl(q_j+\frac{1}{2}k_0\bigr)b_jB_j&=&0,\label{1}\\
\sum_j e^{iq_j\pi}\bigl(q_j-\frac{1}{2}q_0\bigr)a_jB_j&=&0,\label{2}\\
\sum_j a_jB_j&=&0,\label{3}\\
\sum_j e^{iq_j\pi}b_jB_j&=&0,\label{4}
\end{eqnarray}
for $B_j$.  
Another 4 equations determine the transport coefficients, 
\begin{eqnarray}
r&=&-1+\frac{1}{2}\sum_j A_ja_j, \label{5}\\
\rho&=&\frac{1}{2}\sum_j B_jb_j, \label{6}\\
t&=&-\frac{1}{2}\sum_j B_j e^{iq_j\pi}q_ja_j, \label{7}\\
\eta&=&\frac{1}{2}\sum_j A_j e^{iq_j\pi}q_jb_j. \label{8}
\end{eqnarray}
In the above equations, the index $j$ runs over $j=1, 2, 3, 4$.
\par
From Eqs.(\ref{1}) to (\ref{4}), we find immediately
\begin{equation}
B_j=0, \quad j=1,2,3,4,
\end{equation}
namely, an antisymmetric current (circular current) is not induced in the ring. Then from Eqs. (\ref{6}) and (\ref{7}), 
we find $\rho=0$ and $t=0$. This means that if the incident spin is up, the spin-flip transmission and reflection are 
strictly forbidden. Specifically, if the incident energy of the up-spin is below threshold $-\mu<E<\mu$, the transmission 
to the right domain is totally forbidden. 
\par
In Figs. \ref{Double_weak} and \ref{Double_strong}, the numerical results of the transmission spectrum $T_{\uparrow\rightarrow\uparrow}$ are shown for weak and intermediate coupling cases, and for strong coupling 
case. The spin-conserved reflection $R_{\uparrow\rightarrow\uparrow}$ are given by 
$$
R_{\uparrow\rightarrow\uparrow}=1-T_{\uparrow\rightarrow\uparrow}.
$$
As shown in the figures, the transmission spectra have structures of repeated peaks and sharp dips. The peak of 
the transmission are due to the resonant transmission by the eigenstates shown in Fig. \ref{Epermu}. 
The value of $E$ at sharp zero points are given by
$E=n^2, \quad n=1,2,3,\cdots$ 
numerically exactly. This dips and zeros are attributed to the destructive interference due to the two-resonant 
transmissions. 
This ordered structure is disturbed in the strong coupling case shown in Fig. \ref{Double_strong}, at about 
$E=45$ in $\mu=10$ and $E=55$ in $\mu=15$. The origin of this disorder may be assigned to the degeneracy 
of the eigenstates with opposite chirality as shown in Fig. \ref{Epermu}. Comparing these results 
with those for a single-path case shown in Fig. \ref{SingleWeak}, we can see how the quantum path interference 
in the Berry ring changes dramatically the structure of transmission and reflection of a traveling spin. It is noticeable 
that, if one neglects the fine interference structures, the gross features of the line-shape of 
$T_{\uparrow\rightarrow\uparrow}$ for the double-paths are in agreement with those for a single-path case as 
can be seen comparing Fig. 5 and Fig. 8. 

\begin{figure}
\centering
\includegraphics[scale = 0.6]{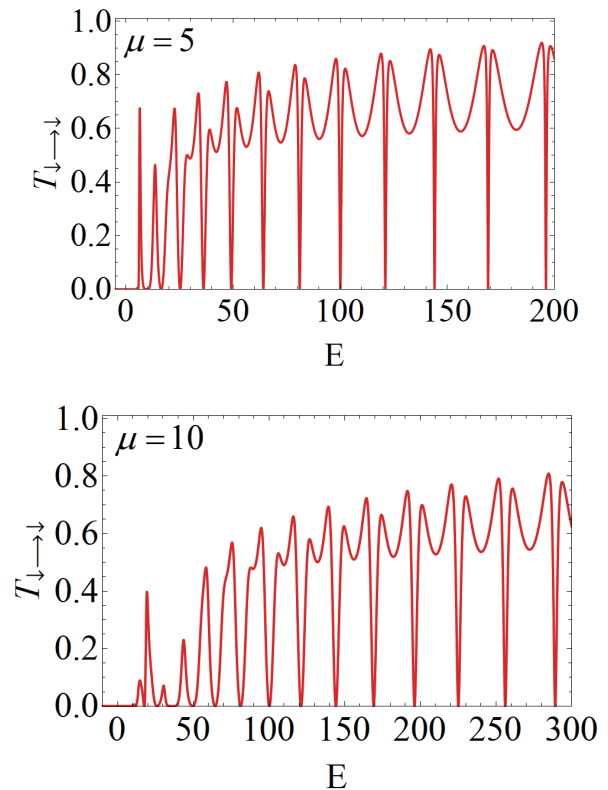}
\caption{Spin-conserved transmission spectrum for the incident minority down spin in the Berry ring 
with $\mu=5$ and $\mu=10$. } 
\label{Double_downspin}
\end{figure}
In Fig.\ref{Double_downspin}, two examples of the transmission spectra for the incident minority spin, namely 
$T_{\downarrow\rightarrow\downarrow}$ are shown. The features of resonant transmission and anti-resonant 
dips are qualitatively the same as those for the incident up-spin case. Specifically, the energies of 
zero point of transmission are the same, $E=n^2$. This means that at this special values of energy, 
the Berry ring, or the double domain walls in ferromagnets work as a complete insulator of electrons. 

\section{Conclusion}
In the present paper, we studied theoretically the transmission of an electron with spin through domain walls in 
ferromagnetic materials. The one-dimensional scattering problem through a domain-wall was formulated as a transport 
problem in transmission lines. Especially, the coherent transmission through a pair of domain walls with opposite chirality 
of the magnetic moment is clarified by a simplified model of one-dimensional line containing a ring with Berry phase.  
For a closed circuit, an analogy of the orbital angular momentum and the emergent spin-orbit interaction was introduced 
by an $SU(2)$ gauge transformation.  
It was shown that the quantum-path interference due to the Berry phase effect greatly modifies the transmission 
spectrum of the electron from that of a single-path setting. 
\par
In experiments, the Fermi energy of the electrons will be a variable external parameter by changing the applied 
voltages. Then spin-dependent drastic changes of the transmission amplitude as shown in Figs. \ref{Double_weak} and 
\ref{Double_strong} will be a signature of the Berry phase effect.
\par
It was shown that, unlike the Aharonov-Bohm ring, persistent current does not exist in the Berry ring because of 
the time-reversal symmetry. 
The coexistence of Berry phase effect and the Aharonov-Bohm effect will be an interesting subject in future. 
It will be worthwhile to clarify the effect of the magnetic flux linking the Berry ring.  
\par
In the present paper, it is assumed that the transport of the electron is ballistic and all the process is coherent. 
In actual experimental settings, the existence of impurities and 
disorders will be inevitable. Since the sharp resonance and the anti-resonances in the transmission spectrum is a 
result of the interference effect in the double paths, the inelastic scattering will have a tendency to destroy 
the structures. The effect of the elastic scatterings by the disorder, on the other hand, will pose interesting 
problems such as the weak localization and universal conductance fluctuations in the presence of Berry phase effect.

\par
We thank Dr. J. Inoue for valuable comments on the symmetry properties of this model. One of the authors (Y. K) 
thanks Prof. H. Katayama-Yoshida for drawing our attention to the experimental works on the domain wall in 
ferromagnetic nanowires. 
\par
This work was supported by JSPS KAKENHI Grants NO. 19K03696.



\end{document}